\documentclass[aps,prd,preprintnumbers,showpacs,superscriptaddress,nofootinbib,amsmath,amssymb,floats,floatfix,showkeys,notitlepage,longbibliography]{revtex4-1}

\usepackage{graphicx,epsf}
\usepackage{mwe,tikz}
\usepackage{orcidlink}
\usepackage{comment}
\usepackage{graphicx}
\usepackage{subfigure}
\usepackage{sans}
\usepackage{hyperref}
\hypersetup{colorlinks=true,linkcolor=blue,urlcolor=blue,citecolor=blue}
\usepackage[toc,page]{appendix}
\usepackage[normalem]{ulem}
\usepackage{adjustbox}
\usepackage{latexsym}
\usepackage{amsmath}
\usepackage{amssymb}
\usepackage{amsfonts}
\usepackage{dcolumn}
\usepackage{bm}
\usepackage{tikz}
\usepackage{bigints}
\usepackage{array,tabularx,multirow,booktabs}
\usepackage[tracking=true]{microtype}
\SetTracking{}{500}
\SetTracking{encoding={*}, shape=sc}{40}
\UseRawInputEncoding 
\allowdisplaybreaks

\newcommand{\ve}{\varepsilon}
\newcommand{\be}{\begin{eqnarray}}
\newcommand{\ee}{\end{eqnarray}}
\newcommand{\bea}{\begin{eqnarray}}
\newcommand{\eea}{\end{eqnarray}}

\def\comment#1{}

\newcommand{\om}{\tilde{\omega}}

\definecolor{darkred}{rgb}{.8,0,0}

\definecolor{darkblue}{rgb}{0,0,.7}

\definecolor{darkgreen}{rgb}{0,.7,0}

\def\be{\begin{equation}}
\def\ee{\end{equation}}
\def\bestar{\begin{equation*}}
\def\eestar{\end{equation*}}

\newcommand{\brr}{\begin{array}}\newcommand{\err}{\end{array}}
\newcommand{\bit}{\begin{itemize}}\newcommand{\eit}{\end{itemize}}
\newcommand{\ben}{\begin{enumerate}}\newcommand{\een}{\end{enumerate}}

\newcommand{\ba}{\begin{array}}
\newcommand{\ea}{\end{array}}

\graphicspath{{./images/} }

\newcolumntype{M}[1]{>{\centering\arraybackslash}m{#1}}
\newcolumntype{N}{@{}m{0pt}@{}}

\newcounter{sxn}

\newcounter{axn}

\newdimen\mybaselineskip
\newcommand{\beeq}{\begin{equation}}
\newcommand{\eneq}{\end{equation}}
\newcommand{\beqn}{\begin{eqnarray}}
\newcommand{\eeqn}{\end{eqnarray}}

\newcommand{\beal}{\setcounter{letter}{1} \begin{eqnarray}}
\newcommand{\eeal}{\addtocounter{equation}{1} \end{eqnarray}}

\newcommand{\larrow}{\,\,\,\,\hbox to 30pt{\rightarrowfill}
\,\,\,\,}
\newcommand{\slarrow}{\,\,\,\hbox to 20pt{\rightarrowfill}
\,\,\,}

\def\la{\raise.16ex\hbox{$\langle$}\lower.16ex\hbox{}  }
\def\ra{\, \raise.16ex\hbox{$\rangle$}\lower.16ex\hbox{} }

\def\psibar{ \psi \kern-.65em\raise.6em\hbox{$-$} \lower.6em\hbox{} }
\def\psibarb{ \psi \kern-.65em\raise.6em\hbox{$-$}  }

\begin{document} \sloppy


\title{Investigating the Connection between Generalized Uncertainty Principle and Asymptotically Safe Gravity in Black Hole Signatures through Shadow and Quasinormal Modes}

\author{Gaetano Lambiase
\orcidlink{0000-0001-7574-2330}}
\email{lambiase@sa.infn.it}
\affiliation{Dipartimento di Fisica ``E.R Caianiello'', Università degli Studi di Salerno, Via Giovanni Paolo II, 132 - 84084 Fisciano (SA), Italy.}
\affiliation{Istituto Nazionale di Fisica Nucleare - Gruppo Collegato di Salerno - Sezione di Napoli, Via Giovanni Paolo II, 132 - 84084 Fisciano (SA), Italy.}

\author{Reggie C. Pantig
\orcidlink{0000-0002-3101-8591}}
\email{rcpantig@mapua.edu.ph}
\affiliation{Physics Department, Map\'ua University, 658 Muralla St., Intramuros, Manila 1002, Philippines}

\author{Dhruba Jyoti Gogoi \orcidlink{0000-0002-4776-8506}}
\email{moloydhruba@yahoo.in}
\affiliation{Department of Physics, Dibrugarh University,
Dibrugarh 786004, Assam, India.}

\author{Ali \"Ovg\"un
\orcidlink{0000-0002-9889-342X}
}
\email{ali.ovgun@emu.edu.tr}
\affiliation{Physics Department, Eastern Mediterranean University, Famagusta, 99628 North Cyprus via Mersin 10, Turkey.}

\begin{abstract}
The links between the deformation parameter $\beta$ of the generalized uncertainty principle (GUP) to the two free parameters $\hat{\omega}$ and $\gamma$ of the running Newtonian coupling constant of the Asymptotic Safe gravity (ASG) program, has been conducted recently in [Phys.Rev.D 105 (2022) 12, 124054]. 
In this paper,
we test these findings by calculating and examining the shadow and quasinormal modes of black holes,
and demonstrate that the approach provides a theoretical framework for exploring the interplay between quantum gravity and GUP.
Our results confirm the consistency of ASG and GUP, and offer new insights into the nature of black holes and their signatures. The implications of these findings for future studies in quantum gravity are also discussed.
\end{abstract}

\date{\today}


\maketitle
\section{Introduction} \label{sec1}
Quantum gravity is a field of theoretical physics that attempts to describe the behavior of gravity at the quantum level. Although Einstein's general relativity has made remarkable contributions to our understanding of gravity, the existence of singularities and instabilities within classical black holes suggests that a more fundamental explanation is required to replace the classical framework.  Quantum gravity aims to reconcile two fundamental theories of physics: quantum mechanics and general relativity \cite{Burgess:2003jk,Capozziello:2011et}. Several approaches to quantum gravity have been proposed, such as loop quantum gravity, string theory, and asymptotic safet gravity (ASG), among others \cite{Ashtekar:2004vs,Rovelli:1997yv,Aharony:1999ti,Bonanno:2000ep,Bonanno:2001xi,Reuter:2003ca}.

One of the most crucial consistency requirements among various constraints is the ability to recover a gravitational effective field theory in the infrared (IR) by starting from the deep ultraviolet (UV). However, only a few theories have met this criterion so far. Asymptotically safe gravity \cite{Bonanno:2000ep,Bonanno:2001xi,Reuter:2003ca} is one such theory that has emerged as a minimal yet promising proposal. It suggests that quantum gravity can be described by a quantum field theory (QFT) whose UV behavior is governed by an interacting fixed point of the gravitational renormalization group (RG) flow. This fixed point acts as an attractor for a subset of RG trajectories, providing a UV completion for the theory and making it renormalizable according to Wilson's approach. The RG improvement has proven to be a valuable tool for investigating the possible implications of asymptotically safe gravity in cosmology and astrophysics, particularly during the emergence of asymptotically safe phenomenology. This technique has even led to the development of a new program, which is now separate from asymptotically safe gravity and known as "scale-dependent gravity". However, more thorough and rigorous derivations and arguments based on either functional integrals or effective action are necessary to understand the modifications to classical black holes and early-universe cosmology induced by asymptotic safe gravity. These fundamental approaches have been the primary focus of the asymptotic safe gravity studies in recent years \cite{Koch:2016uso,Bonanno:2002zb,Fathi:2019jid,Contreras:2017eza,Contreras:2018dhs,Rincon:2018lyd,Rincon:2018dsq,Rincon:2019cix,Contreras:2018gpl,Rincon:2018sgd,Rincon:2017goj}.

On the other hand, a significant area of research aimed at describing the relationship between quantum effects and gravity is referred to as the Generalized Uncertainty Principle (GUP) which has focused on how the Heisenberg Uncertainty Principle (HUP) should be modified when accounting for gravity. Because gravity plays a central role in these investigations, the most relevant modifications to the HUP have been proposed in string theory, loop quantum gravity, deformed special relativity, and studies of black hole physics \cite{Maggiore:1993rv,Kempf:1994su}. The dimensionless deforming parameter of the GUP, denoted by $\beta$, is not fixed by the theory, although it is typically assumed to be on the order of one (as in some models of string theory, for example, Ref. \cite{Scardigli:1999jh,Adler:1999bu,Capozziello:1999wx,Scardigli:2003kr,Ovgun:2017hje,Ovgun:2016roz,Ovgun:2015box,Ovgun:2015jna,Ali:2009zq,Chen:2014jwq,Tawfik:2015rva,Casadio:2013aua}).

The key attribute that defines a black hole is its event horizon, which marks the point beyond which particles cannot escape. This immense gravitational pull traps all physical particles, including light, inside the event horizon. In contrast, outside this boundary, light can escape \cite{Synge:1966okc}. The matter that surrounds a black hole and is pulled inward is known as accretion. Over time, the accretion becomes heated due to viscous dissipation and emits bright radiation at various frequencies, including radio waves that can be detected by radio telescopes. The accreting material creates a bright background with a dark area over it, known as the black hole shadow \cite{Luminet:1979nyg}. Although the concept of the black hole shadow has been around since the 1970s, it wasn't until Falcke et al. \cite{Falcke:1999pj} that the idea of imaging the black hole shadow at the center of our Milky Way was first proposed. The Event Horizon Telescope has recently captured the image of the black hole shadow in the Messier 87 galaxy and Sagittarius A* \cite{EventHorizonTelescope:2019dse,EventHorizonTelescope:2022xnr}. As a result, the black hole shadow has become a popular subject in today's literature since the shadow can be used to extract information about the deviations in the spacetime geometry \cite{Ovgun:2018tua,Ovgun:2020gjz,Ovgun:2019jdo,Kuang:2022xjp,Kumaran:2022soh,Mustafa:2022xod,Cimdiker:2021cpz,Okyay:2021nnh,Atamurotov:2022knb,Pantig:2022qak,Abdikamalov:2019ztb,Abdujabbarov:2016efm,Atamurotov:2015nra,Papnoi:2014aaa,Abdujabbarov:2012bn,Atamurotov:2013sca,Cunha:2018acu,Gralla:2019xty,Belhaj:2020okh,Belhaj:2020rdb,Konoplya2019,Wei2019,Ling:2021vgk,Kumar:2020hgm,Kumar2017EPJC,Cunha:2016wzk,Cunha:2016bpi,Cunha:2016bjh,Zakharov:2014lqa,Tsukamoto:2017fxq,Chakhchi:2022fls,Li2020,EventHorizonTelescope:2021dqv,Vagnozzi:2022moj}. These deviations might be the cause of some parameters from various alternative theories of gravity \cite{Pantig:2022ely,Pantig:2022gih,Lobos:2022jsz,Uniyal:2022vdu,Ovgun:2023ego,Rayimbaev:2022hca,Uniyal:2023inx,Panotopoulos:2021tkk,Panotopoulos:2022bky,Khodadi:2022pqh,Khodadi:2021gbc,Zhao:2023uam}, or the astrophysical environment where the black hole is immersed in \cite{Pantig:2022whj,Pantig:2022sjb, Pantig:2023yer, Wang:2019skw, Roy:2020dyy,Xu:2018wow,Konoplya:2021ube,Konoplya:2022hbl,Anjum:2023axh}. In this paper, we also aim to find constraints of the GUP parameters $\hat{\omega}$ and $\gamma$, by extending the formalism in Refs. \cite{Perlick:2015vta, Perlick:2021aok} with the shadow radius instead of the angular radius. We also analyze the behavior of the black hole shadow using these constraints.

Black holes are intriguing objects in the Universe that are closely linked to the production of gravitational waves. Quasinormal modes are a fascinating feature of black hole physics that describe the damped oscillations of a black hole, characterized by complex frequencies \cite{Andersson:1996cm,Andersson:2003fh,Ferrari:2007dd,Berti:2009kk,Kokkotas:1999bd,Nollert:1999ji,Boudet:2022wmb,Berti:2022xfj,Cardoso:2016rao,Berti:2015itd}. These modes are important because they provide valuable information about the properties of black holes, such as their mass, angular momentum, and the nature of surrounding spacetime. The study of quasinormal modes is crucial for understanding the structure and evolution of black holes, and their role in astrophysical phenomena such as gravitational wave signals. Recent research has extensively explored the properties of gravitational waves and quasinormal modes of black holes in various modified gravity theories \cite{ 
Ovgun:2017dvs,Bouhmadi-Lopez:2020oia,Gogoi:2022wyv,Gogoi:2021dkr,Gogoi:2021cbp,Gogoi:2022ove,Gogoi:2023kjt,Gundlach:1993tn,Schutz:1985km,Iyer:1986np,Konoplya:2003ii,Daghigh:2011ty,Daghigh:2008jz,Zhidenko:2003wq,Zhidenko:2005mv,Lepe:2004kv,Chabab:2016cem,Chabab:2017knz,Konoplya:2011qq,Hatsuda:2019eoj,Eniceicu:2019npi,Gonzalez:2021vwp,Rincon:2021gwd,Panotopoulos:2020mii,Panotopoulos:2019qjk,Rincon:2018sgd,Gonzalez:2022ote,Yang:2022xxh,Yang:2022ifo,Ovgun:2019yor}.

The main aim of this paper is to investigate the ASG parameter through the analysis of the quasinormal modes and shadow of a black hole. Additionally, we aim to establish a correlation between the unconstrained variables of ASG, specifically the renormalization scale, and the deforming parameter $\beta$ of the generalized uncertainty principle (GUP).

We program the paper as follows: In Sect. \ref{sec2}, we briefly review the black hole in asymptotically safe gravity. In Sect. \ref{sec3}, we study the black hole shadow by initially finding the constraints, and examining how the shadow behaves relative to some observer. Then, in Sect. \ref{sec4}, we study the scalar and 
electromagnetic perturbation and associated quasinormal modes, then we investigate the time evolution profiles of the perturbations and the quasinormal modes generated by such a black hole. We then form conclusive remarks in Sect. \ref{Conc}. Finally, we give research directions. In this paper, we use the metric signature $(-,+,+,+)$. 

\section{Black hole in Asymptotically safe gravity} \label{sec2}
The static and spherically symmetric metric by ASG improved Schwarzschild black hole can be expressed \cite{Bonanno:2000ep}
by 
\be
\mathrm{d}s^2 = -f(r)\mathrm{d} t^2 + f(r)^{-1} \mathrm{d} r^2 + C(r) (\mathrm{d \theta^2 + sin^2 \theta \: d \phi^2}),
\label{metric}
\ee
where the lapse function $f(r)$ is 
\be
\label{ASGmetric}
f(r) = 1 - \frac{2 M G(r)}{r}=1-\frac{2 G_0 M r^2}{r^3 +  \om G_0 \hbar (r + \gamma G_0 M)}\,,
\ee
with $G(r)=\frac{G_0 r^3}{r^3+\tilde{\omega} G_0 \hbar\left(r+\gamma G_0 M\right)}$ where $G_0$ is  Newton constant. $\om$ and $\gamma$ are dimensionless numerical
parameters and $M$ the mass of the black hole. For $\om = 0$ we recover the standard Schwarzschild metric.

In Ref. \cite{Lambiase:2022xde}, it has been shown that there is a link between the deformation parameter $\beta$ of the generalized uncertainty principle (GUP) to the two free parameters $\omega$ and $\gamma$ of the running Newtonian coupling constant of the Asymptotic Safe gravity (ASG) program. In order to proceed, we express Eq. \eqref{ASGmetric} as a small perturbation around the Schwarzschild metric \cite{Lambiase:2022xde}:
\be
f(r) = 1 - \frac{2G_0 M}{r} + \frac{2 \,G_0^2 \,\om \,M \hbar \,(r + \gamma G_0 M)}{r^4} + {\cal O}(\frac{1}{r^5})
= 1 - \frac{2G_0 M}{r} + \ve(r)
\label{ve}
\ee
with
\be
\ve(r) \simeq \frac{2 \,G_0^2 \,\om \,M \hbar \,(r + \gamma G_0 M)}{r^4}\,,
\label{ver}
\ee
Note that $|\ve(r)|\ll2G_0 M/r$ for any $r>2G_0M$.
The horizon of the black hole is 

\be
r_h = a - \frac{a\ve(a)}{1+\ve(a)+a\ve'(a)} = a\left[1-\ve(a) + {\cal O}(\ve^2)\right]\,,
\label{rh}
\ee
where $a=2G_0M$.

To establish a connection between the Asymptotic Safe Gravity (ASG) and the Generalized Uncertainty Principle (GUP), one may compare the first orders of the expansions of the GUP-deformed Hawking temperature and the ASG-Schwarzschild temperature \cite{Lambiase:2022xde}. This allows us to obtain the relationship between the deformation parameter $\beta$ and the two free parameters $\om$ and $\gamma$ 

\be
\label{beta}
\beta \ = \ -\,4\, \pi^2 \, \om \, (1+\gamma).
\ee

Then, using the ASG-improved Newtonian potential, the parameter $\om$ is fixed to 
\be
\om = - \frac{41}{10\pi}\,.
\label{om}
\ee
It is worth noting that the ASG parameter $\gamma$ is not a fixed value, although classical general relativistic arguments do set $\gamma = 9/2$. With this value of $\gamma$, the $\beta$ parameter of the GUP can be determined to be approximately \be\beta=\frac{451 \pi}{5},\ee 
which is of the order of $10^2$ as predicted by certain string theory models.

\section{Constraints using the black hole shadow} \label{sec3}
Here, we will first find constraints to $\gamma$ by using the empirical data provided by the EHT collaboration for Sgr. A* and M87*, which is summarized in Table \ref{tab1}.§
\begin{table} [!ht]
    \centering
    \begin{tabular}{ p{2cm} p{3.5cm} p{4.5cm} p{2cm}}
    \hline
    \hline
    Black hole & Mass $m$ ($M_\odot$) & Angular diameter: $2\theta_\text{sh}$ ($\mu$as) & Distance (kpc) \\
    \hline
    Sgr. A*   & $4.3 \pm 0.013$x$10^6$ (VLTI)    & $48.7 \pm 7$ (EHT) &   $8.277 \pm 0.033$ \\
    M87* &   $6.5 \pm 0.90$x$10^9$  & $42 \pm 3$   & $16800$ \\
    \hline
    \end{tabular}
    \caption{Black hole observational constraints.}
    \label{tab1}
\end{table}
Also for convenience, we will only consider a constant polar angle $\theta = \pi/2$ in the analysis of null orbits. Using the black hole metric in Eq. \eqref{metric} with the lapse function $f(r)$ in Eqs. \eqref{ve}-\eqref{ver}, the null geodesics along the equatorial plane  can be derived through the Lagrangian
\begin{equation}
    \mathcal{L} = \frac{1}{2}\left( -A(r) \dot{t} +B(r) \dot{r} + C(r) \dot{\phi} \right).
\end{equation}
Applying the variational principle, the two constants of motion can be obtained
\begin{equation} \label{econs}
    E = A(r)\frac{dt}{d\lambda}, \quad L = C(r)\frac{d\phi}{d\lambda},
\end{equation}
whereas the impact parameter, which is important in orbital motion analysis, is defined as
\begin{equation}
    b \equiv \frac{L}{E} = \frac{C(r)}{A(r)}\frac{d\phi}{dt}.
\end{equation}
We can obtain how the radial coordinate changes with the azimuthal angle by setting $ds^2 = 0$. That is,
\begin{equation} \label{eorb}
    \left(\frac{dr}{d\phi}\right)^2 =\frac{C(r)}{B(r)}\left(\frac{h(r)^2}{b^2}-1\right),
\end{equation}
where, by definition \cite{Perlick2015},
\begin{equation}
    h(r)^2 = \frac{C(r)}{A(r)}.
\end{equation}

The simple definition above allows one to obtain the location of the photonsphere radius by taking $h'(r) = 0$, where the prime denotes differentiation with respect to $r$. In doing so, we obtained:
\begin{equation} \label{erps}
    \varepsilon'(r) r^{2}+6 G_0 M -2 \varepsilon(r) r -2 r = 0.
\end{equation}
We can only do a numerical analysis in obtaining the photonsphere radius $r_\text{ps}$, which, by the above equation, is affected by the parameters $\tilde{\omega}$ and $\gamma$. This is important since the shadow cast, and how the shadow radius behaves, depend on the photonsphere radius as the critical impact parameter is evaluated in $r_\text{ps}$.

If a certain observer is located at the coordinates ($(t_\text{obs},r_\text{obs},\theta_\text{obs} = \pi/2, \phi_\text{obs} = 0)$), then one can construct \cite{Perlick:2018} the relation
\begin{equation}
    \tan(\alpha_{\text{sh}}) = \lim_{\Delta x \to 0}\frac{\Delta y}{\Delta x} = \left(\frac{C(r)}{B(r)}\right)^{1/2} \frac{d\phi}{dr} \bigg|_{r=r_\text{obs}},
\end{equation}
which can be alternatively expressed as
\begin{equation} \label{eangrad}
    \sin^{2}(\alpha_\text{sh}) = \frac{b_\text{crit}^{2}}{h(r_\text{obs})^{2}},
\end{equation}
with the help of Eq. \eqref{eorb}. In general, since some spacetime metrics do not have identical expressions for $h(r)$, the expression for the critical impact parameter is given by \cite{Pantig:2022sjb, Pantig:2022ely}
\begin{equation} \label{ebcrit}
    b_\text{crit}^2 = \frac{h(r_\text{ps})}{\left[B'(r_\text{ps})C(r_\text{ps})-B(r_\text{ps})C'(r_\text{ps})\right]} \Bigg[h(r_\text{ps})B'(r_\text{ps})C(r_\text{ps})-h(r_\text{ps})B(r_\text{ps})C'(r_\text{ps})-2h'(r_\text{ps})B(r_\text{ps})C(r_\text{ps}) \Bigg].
\end{equation}
Using the lapse function $f(r)$ in Eqs. \eqref{ve}-\eqref{ver}, and Eq. \eqref{erps},
\begin{equation} \label{ebcrit2}
	b_\text{crit}^2= \frac{4 r_\text{ps}^3}{\varepsilon'(r_\text{ps}) r_\text{ps}^{2}-2 G_0 M +2 \varepsilon(r_\text{ps}) r_\text{ps} +2 r_\text{ps}}.
\end{equation}
Finally, the shadow radius can be sought-off as
\begin{equation} \label{e32}
	R_\text{sh} = b_\text{crit}\sqrt{1 - \frac{2G_0 M}{r_\text{obs}} + \ve(r_\text{obs})}.
\end{equation}

By considering the distance $D$ of the SMBH from the galactic center, the classical shadow diameter can be found through the standard arclength equation
\begin{equation} \label{earc}
    d_\text{sh} = \frac{D \theta_\text{sh}}{M}.
\end{equation}
The calculated values for the diameter of M87* and Sgr. A*'s shadows are as follows:  These are $d^\text{M87*}_\text{sh} = (11 \pm 1.5)M$, and $d^\text{Sgr. A*}_\text{sh} = (9.5 \pm 1.4)M$, respectively \cite{EventHorizonTelescope:2019dse, EventHorizonTelescope:2022xnr}. Note that sometimes, it is also useful to use the Schwarzschild deviation parameter $\lambda$ to find constraints to parameters of a certain BH model. Here, we have used the uncertainties reported in Refs. \cite{EventHorizonTelescope:2021dqv,Vagnozzi:2022moj} to find constraints to $\tilde{\omega}$ and $\gamma$. Note that these uncertainties are tighter than if we used Eq. \eqref{earc}. The interested reader is directed to such reference to see how the $1\sigma$ and $2\sigma$ confidence levels were found.

\begin{figure*}[!ht]
    \centering
    \includegraphics[width=0.48\textwidth]{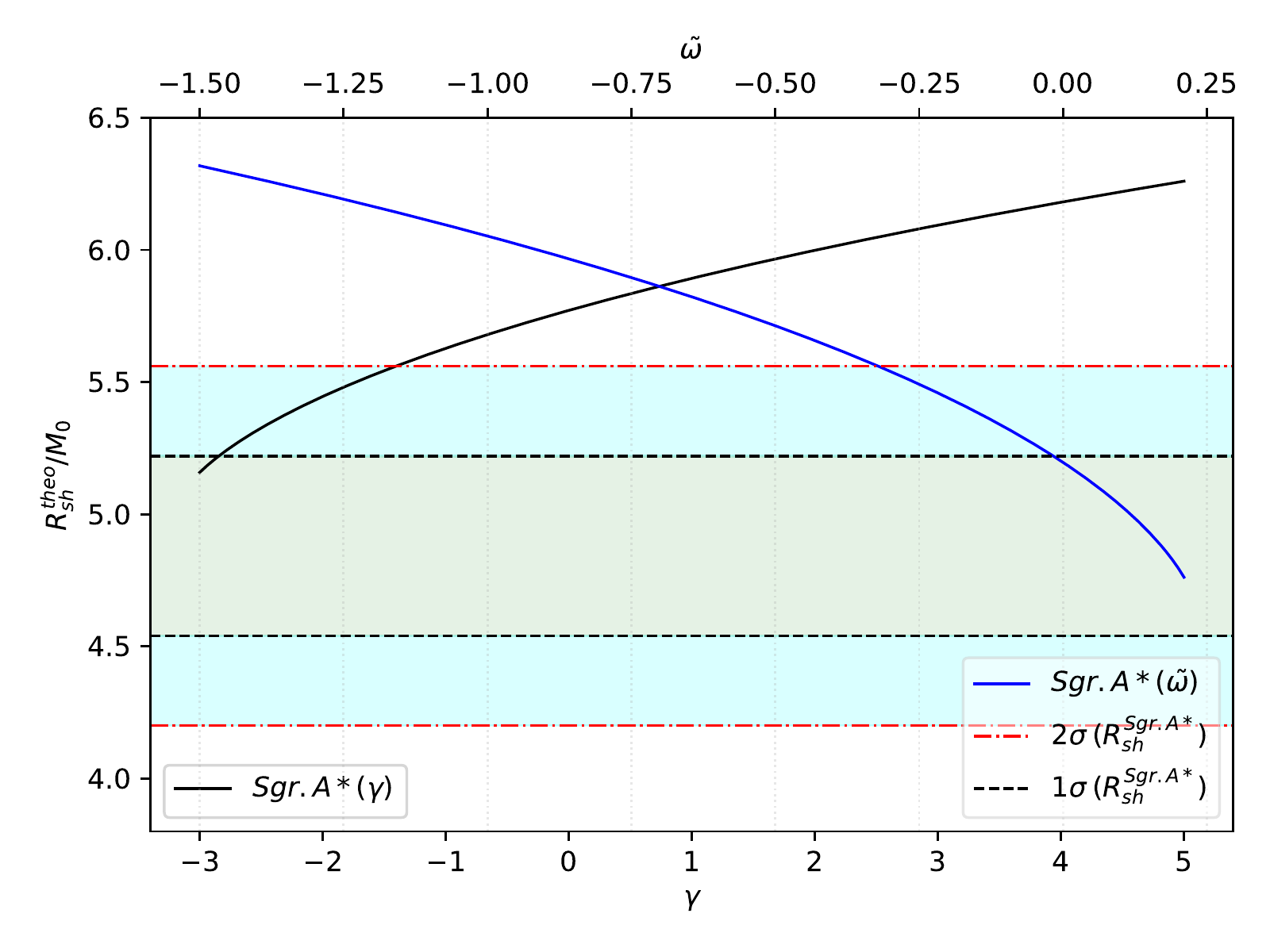}
    \includegraphics[width=0.48\textwidth]{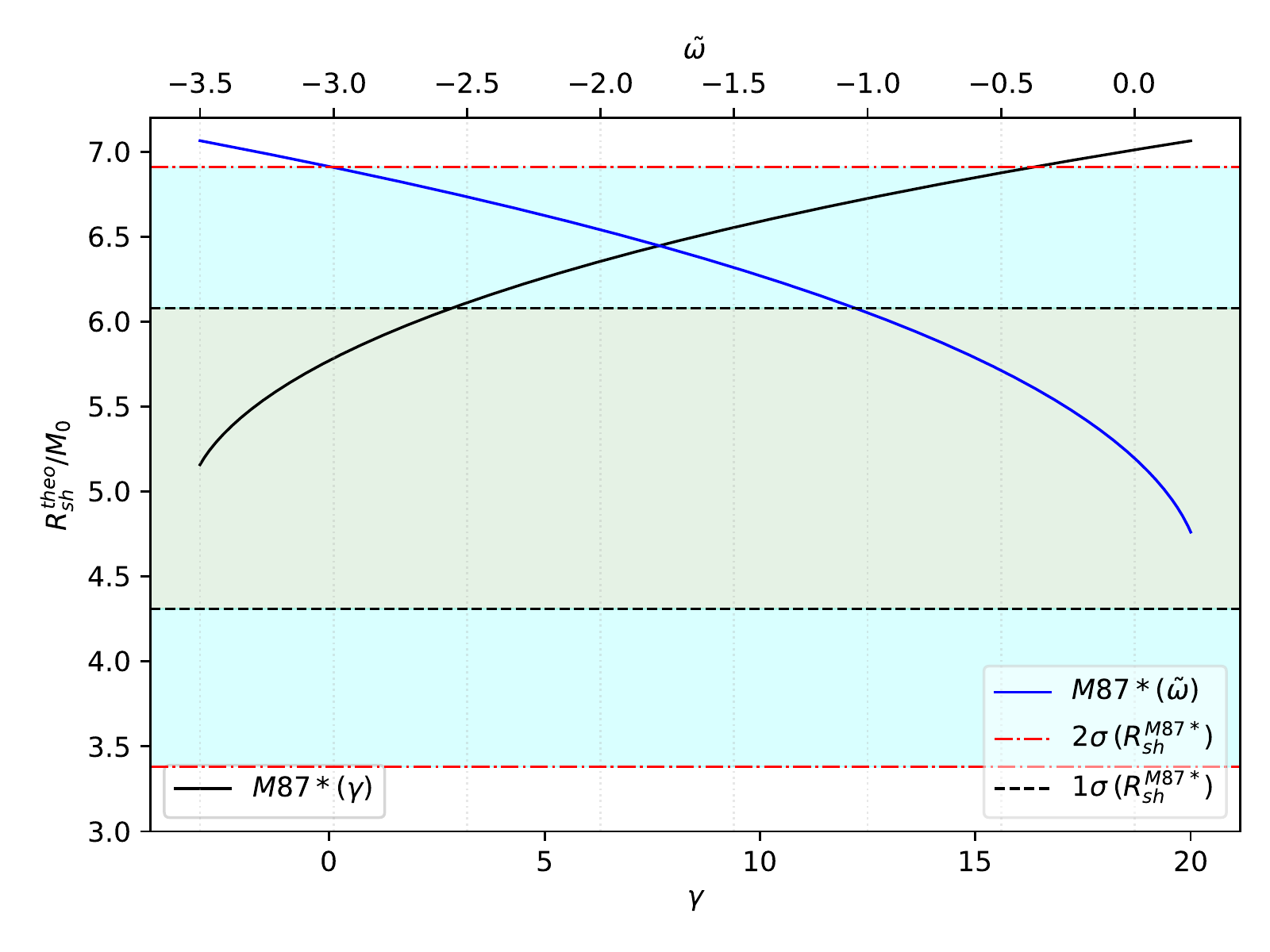}
    \caption{Left: Sgr. A*. Right: M87*. In this plot, the black solid line corresponds to the case of fixed $\om = -41/10\pi$, while the blue solid line is for the case of fixed $\gamma = 9/2$.}
    \label{shacons}
\end{figure*}

Our first constraint plot considers a fixed $\tilde{\omega}$, given by Eq. \eqref{om}. Our results are shown in Fig. \ref{shacons} (black solid lines), and we tabulate the bounds for $\gamma$ in Table \ref{tab2}. In essence, not only that it gave us the relevant bounds for $\gamma$, which in turn gives us the relevant values for $\beta$ in Eq. \eqref{beta}, but it also visualizes how the shadow radius behaves as $\gamma$ (or $\beta$) varies considering a fixed observer given in Table. \ref{tab1}.
\begin{table}[!ht]
    \centering
    \begin{tabular}{clll}
\hline
\hline
{$\gamma$} &  $1\sigma$(upper/lower) & $2\sigma$(upper/lower) \\
\hline
Sgr. A* &        -2.950 / none &       -1.440 / none \\
M87*    &       2.850 / none &       16.36 / none \\
\hline
\end{tabular}
\qquad
\begin{tabular}{clll}
\hline
\hline
{$\beta$} &    $1\sigma$(upper/lower) & $2\sigma$(upper/lower) \\
\hline
Sgr. A* &        -100.468 / none &       -22.67 / none \\
M87*    &       198.360 / none &       894.424 / none \\
\hline
\end{tabular}
\caption{Values of $\gamma$ (left) and the corresponding $\beta$ (right) based on the constraints imposed by the EHT data on the shadow radius (black solid lines in Fig. \ref{shacons}).}
    \label{tab2}
\end{table}
\begin{table}[!ht]
    \centering
    \begin{tabular}{clll}
\hline
\hline
{$\tilde{\omega}$} &    $1\sigma$(upper/lower) & $2\sigma$(upper/lower) \\
\hline
Sgr. A* &        -0.024 / none &       -0.330 / none \\
M87*    &       -1.059 / none &       -3.000 / none \\
\hline
\end{tabular}
\qquad
\begin{tabular}{clll}
\hline
\hline
{$\beta$} &    $1\sigma$(upper/lower) & $2\sigma$(upper/lower) \\
\hline
Sgr. A* &        5.211 / none &       71.653 / none \\
M87*    &       229.942 / none &       651.394 / none \\
\hline
\end{tabular}
\caption{Values of $\tilde{\omega}$ (left) and the corresponding $\beta$ (right) based on the constraints imposed by the EHT data on the shadow radius (blue solid lines in Fig. \ref{shacons}).}
    \label{tab3}
\end{table}
Some models in string theory \cite{Bonanno:2000ep} suggests a fixed value for $\gamma$ equal to $9/2$, which immediately implies a certain value for $\beta$ \cite{Lambiase:2022xde}:
\begin{equation} \label{beta2}
    \beta = \frac{451\pi}{5}.
\end{equation}
It then leads us to constrain $\tilde{\omega}$ to consider string theory for GUP and asymptotically safe gravity, where the results are shown in Fig. \ref{shacons} (blue solid lines). Furthermore, the numerical values of the bounds for $\tilde{\omega}$ and its corresponding value for $\beta$ is in Table \ref{tab3}. Interestingly, we see that the constraint found in M87* for $\beta$ is in good agreement with Eq. \eqref{beta2} as it falls within the range of the uncertainty levels. It can also be concluded that Sgr. A* gives a poor result for such a constraint.

\begin{figure}[!ht]
    \centering
    \includegraphics[width=0.50\textwidth]{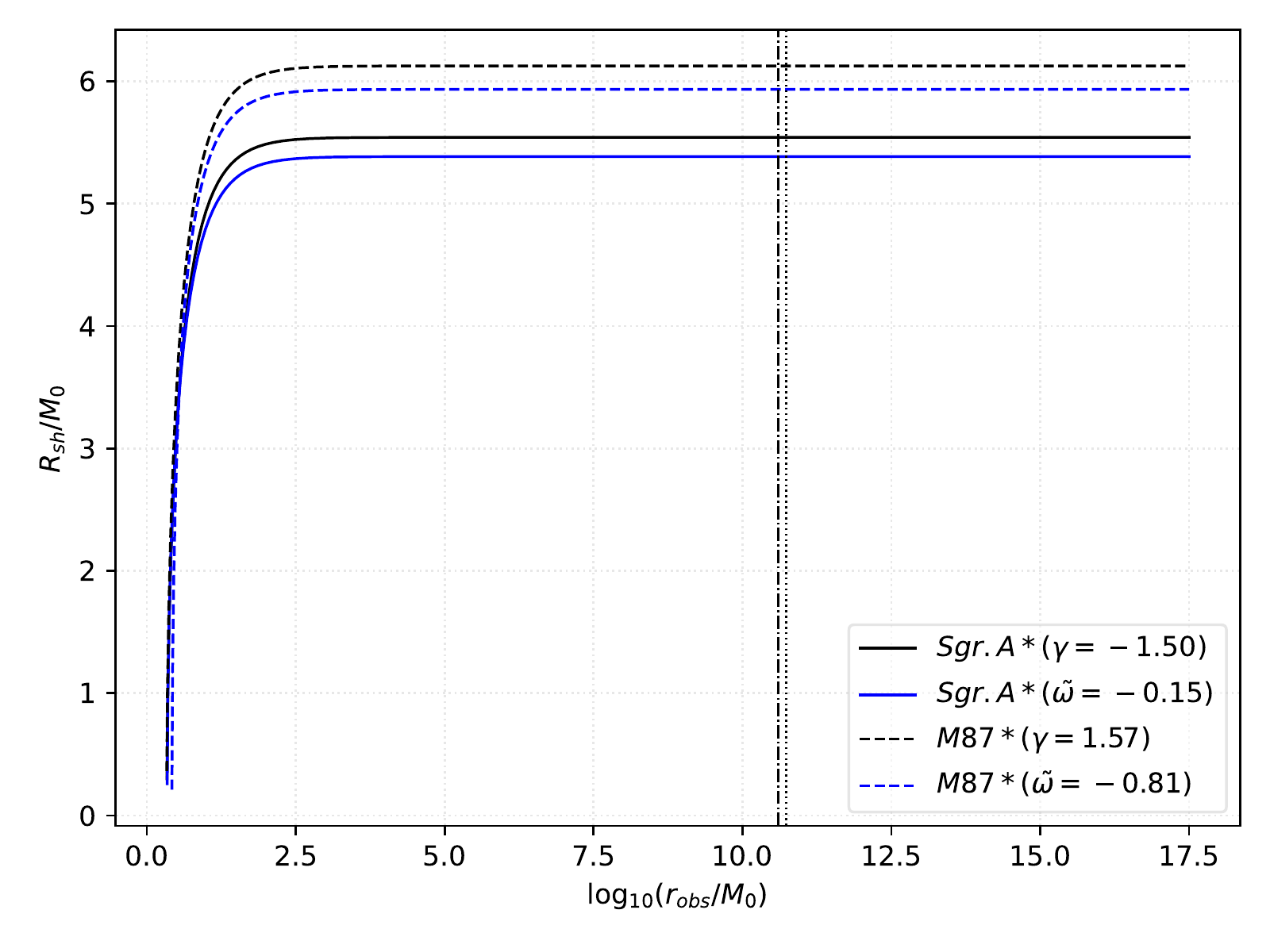}
    \caption{Behavior of the shadow radius as $r_\text{obs}$ varies. The black vertical dotted and dashdot lines correspond to our location from M87* and Sgr. A*, respectively.}
    \label{sharobs}
\end{figure}

Our next aim is to examine the behavior of the shadow radius due to an observer with varying $r_\text{obs}$ for chosen values of the parameters presented in Tables \ref{tab2} and \ref{tab3}. The result in Fig. \ref{sharobs} is simply the plot of the general equation in Eq. \eqref{e32}.
We observe that the effect of the free parameters in asymptotic safe gravity is to increase or decrease the shadow radius at large distances and merely follow the Schwarzschild trend. We observe no peculiar deviation from such a trend. Our results for such constraints indicate, however, that these parameters are best explored through the shadow of M87*.

\section{Quasinormal modes of ASG Black Holes} \label{sec4}
In this section, we study the behaviors of quasinormal modes of the ASG black holes for two different types of perturbations viz. scalar perturbations and electromagnetic perturbations. At first, we derive the associated potentials for scalar and electromagnetic perturbations and then we study the potential behavior with respect to the model parameters of the black hole. The behavior of the potential with the model parameters will provide a rough idea of how the quasinormal modes might vary in the framework. 

In this context, we will presume that the scalar field or electromagnetic field under consideration has a negligible effect on black hole spacetime. To determine the quasinormal modes, we establish Schr\"odinger--like wave equations for each scenario by considering the corresponding conservation relations in the relevant spacetime. The wave equations will take the form of the Klein--Gordon type for scalar fields and the Maxwell equations for electromagnetic fields. We have used the Pad\'e averaged 6th--order WKB approximation method to obtain the quasinormal modes.

By focusing solely on axial perturbation, the perturbed metric can be expressed as follows \cite{Bouhmadi-Lopez:2020oia, Gogoi:2023kjt}:
\begin{equation} \label{pert_metric}
ds^2 = -\, |g_{tt}|\, dt^2 + r^2 \sin^2\!\theta\, (d\phi - p_1(t,r,\theta)\,
dt - p_2(t,r,\theta)\, dr - p_3(t,r,\theta)\, d\theta)^2 + g_{rr}\, dr^2 +
r^2 d\theta^2,
\end{equation}
where the parameters $p_1$, $p_2$, and $p_3$ describe the perturbation affecting the black hole spacetime. The metric functions $g_{tt}$ and $g_{rr}$ represent the zeroth order terms, and they are dependent on $r$ exclusively.

\subsection{Scalar perturbation}
We begin by considering a scalar field with no mass in the vicinity of a previously established black hole. Since we assume that the scalar field has a negligible effect on the spacetime, we can simplify the perturbed metric equation to the following form:
\begin{equation}
ds^2 = -\,|g_{tt}|\, dt^2 + g_{rr}\, dr^2 +r^2 d \Omega^2.
\end{equation}
The properties of the perturbation are characterized by the Klein-Gordon equation associated with the scalar field. In this case, we consider that the scalar field is massless.
Next, we can express the Klein-Gordon equation in curved spacetime for this scenario as follows:
\begin{equation} \label{scalar_KG}
\square \Phi = \dfrac{1}{\sqrt{-g}} \partial_\mu (\sqrt{-g} g^{\mu\nu} \partial_\nu \Phi) = 0.
\end{equation}

This equation explains the quasinormal modes connected with the scalar perturbations which are massless. In the above equation, $\Phi$ represents the associated wave function of the scalar perturbation. It is a function of the coordinates $t, r, \theta$ and $\phi$. We can break down $\Phi$ into spherical harmonics and the radial part which can be represented as:
\begin{equation}
\Phi(t,r,\theta, \phi) = \dfrac{1}{r} \sum_{l,m} \psi_l(t,r) Y_{lm}(\theta, \phi),
\end{equation}
where $l$ and $m$ are the associated indices of the spherical harmonics. The function $\psi_l(t,r)$ is the time-dependent radial wave function. 
Using equation \eqref{scalar_KG}, we can derive the following equation for the scalar perturbation:
\begin{equation}  \label{radial_scalar}
\partial^2_{r_*} \psi(r_*)_l + \omega^2 \psi(r_*)_l = V_s(r) \psi(r_*)_l,
\end{equation}
where $r_*$ is defined as the tortoise coordinate, expressed as:
\begin{equation} \label{tortoise}
\dfrac{dr_*}{dr} = \sqrt{g_{rr}\, |g_{tt}^{-1}|}
\end{equation}

The effective potential $V_s(r)$ in this case, takes on the following explicit form:
\begin{equation} \label{Vs}
V_s(r) = |g_{tt}| \left( \dfrac{l(l+1)}{r^2} +\dfrac{1}{r \sqrt{|g_{tt}| g_{rr}}} \dfrac{d}{dr}\sqrt{|g_{tt}| g_{rr}^{-1}} \right).
\end{equation}
In this equation, the term $l$ represents the multipole moment of the black hole's quasinormal modes.

\subsection{Electromagnetic perturbation}
The next topic is an electromagnetic perturbation, which requires the use of the standard tetrad formalism \cite{Bouhmadi-Lopez:2020oia,Gogoi:2022wyv,Gogoi:2023kjt}. This formalism defines a basis $e^\mu_{a}$ that is related to the black hole metric $g_{\mu\nu}$. The basis satisfies the following conditions:
\begin{align}
e^{(a)}_\mu e^\mu_{(b)} &= \delta^{(a)}_{(b)}  \nonumber \\
e^{(a)}_\mu e^\nu_{(a)} &= \delta^{\nu}_{\mu}  \nonumber \\
e^{(a)}_\mu &= g_{\mu\nu} \eta^{(a)(b)} e^\nu_{(b)}  \nonumber \\
g_{\mu\nu} &= \eta_{(a)(b)}e^{(a)}_\mu e^{(b)}_\nu = e_{(a)\mu} e^{(a)}_\nu.
\end{align}
One can express tensor fields in terms of this basis as shown below:
\begin{align*}
S_\mu &= e^{(a)}_\mu S_{(a)}, \\
S_{(a)} &= e^\mu_{(a)} S_\mu, \\
P_{\mu\nu} &= e^{(a)}_\mu e^{(b)}_\nu P_{(a)(b)}, \\
P_{(a)(b)} &= e^\mu_{(a)} e^\nu_{(b)} P_{\mu\nu}.
\end{align*}

Now, for the electromagnetic perturbation, it is possible to rewrite the Bianchi identity of the field strength $F_{[(a)(b)(c)]} = 0$, in the tetrad formalism as
\begin{align}
\left( r \sqrt{|g_{tt}|}\, F_{(t)(\phi)}\right)_{,r} + r \sqrt{g_{rr}}\,
F_{(\phi)(r), t} &=0,  \label{em1} \\
\left( r \sqrt{|g_{tt}|}\, F_{(t)(\phi)}\sin\theta\right)_{,\theta} + r^2
\sin\theta\, F_{(\phi)(r), t} &=0.  \label{em2}
\end{align}
From the conservation equation, in tetrad formalism, one can further obtain, 
\begin{equation}  \label{em3}
\left( r \sqrt{|g_{tt}|}\, F_{(\phi)(r)}\right)_{,r} + \sqrt{|g_{tt}| g_{rr}}%
\, F_{(\phi)(\theta),\theta} + r \sqrt{g_{rr}}\, F_{(t)(\phi), t} = 0.
\end{equation}
From the time derivative of Eq. %
\eqref{em3} and Eq.s \eqref{em1} and \eqref{em2} one can have the following expression, 
\begin{equation}  \label{em4}
\left[ \sqrt{|g_{tt}| g_{rr}^{-1}} \left( r \sqrt{|g_{tt}|}\, \mathcal{F}
\right)_{,r} \right]_{,r} + \dfrac{|g_{tt}| \sqrt{g_{rr}}}{r} \left( \dfrac{%
\mathcal{F}_{,\theta}}{\sin\theta} \right)_{,\theta}\!\! \sin\theta - r 
\sqrt{g_{rr}}\, \mathcal{F}_{,tt} = 0,
\end{equation}
where $\mathcal{F} = F_{(t)(\phi)} \sin\theta.$ 
Defining $\psi_e \equiv r \sqrt{|g_{tt}|}\, \mathcal{F}$ one can write Eq. \eqref{em4} in the
Schr\"odinger-like form:
\begin{equation}
\partial^2_{r_*} \psi_e + \omega^2 \psi_e = V_e(r) \psi_e,
\end{equation}
where the potential is given by 
\begin{equation}  \label{Ve}
V_e(r) = |g_{tt}|\, \dfrac{l(l+1)}{r^2}.
\end{equation}
This is the potential associated with electromagnetic perturbation. In the next subsection, we shall study the behaviors of these potentials in brief.

\subsection{Behaviour of the potential}
The potential for both types of perturbations depends on the parameters $l$, $\om$, and $\gamma$. From the behavior of the perturbation potential, it is possible to have a preliminary idea of the behavior of quasinormal modes associated with the black hole spacetime.
\begin{figure}[!ht]
    \centering {
    \includegraphics[scale= 0.70]{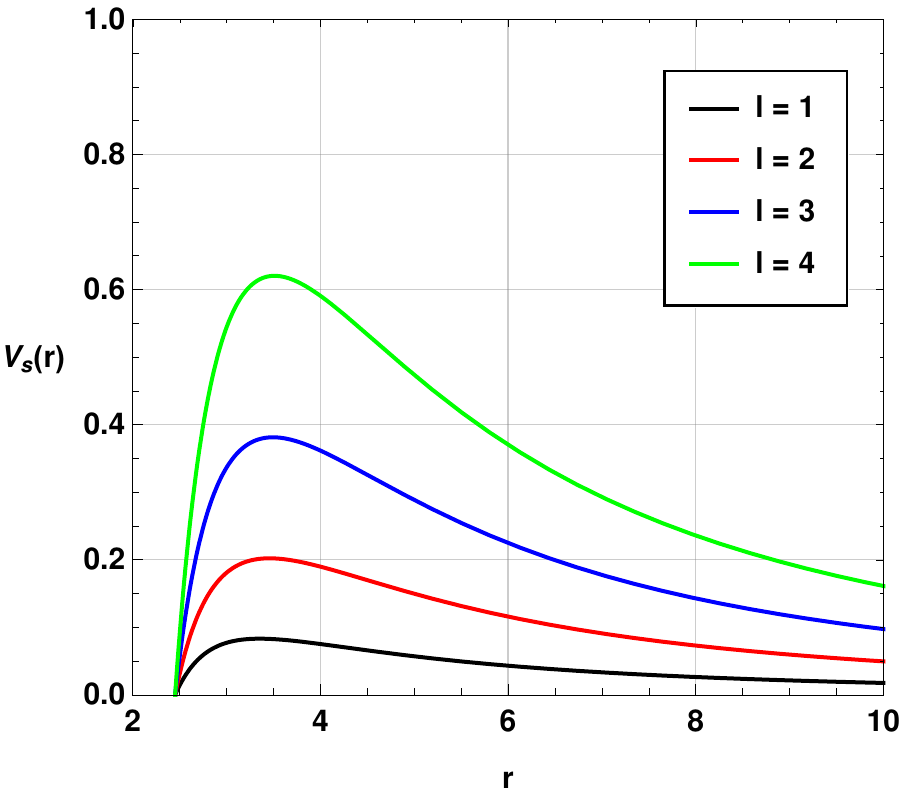} \hspace{1cm}
     \includegraphics[scale=0.710]{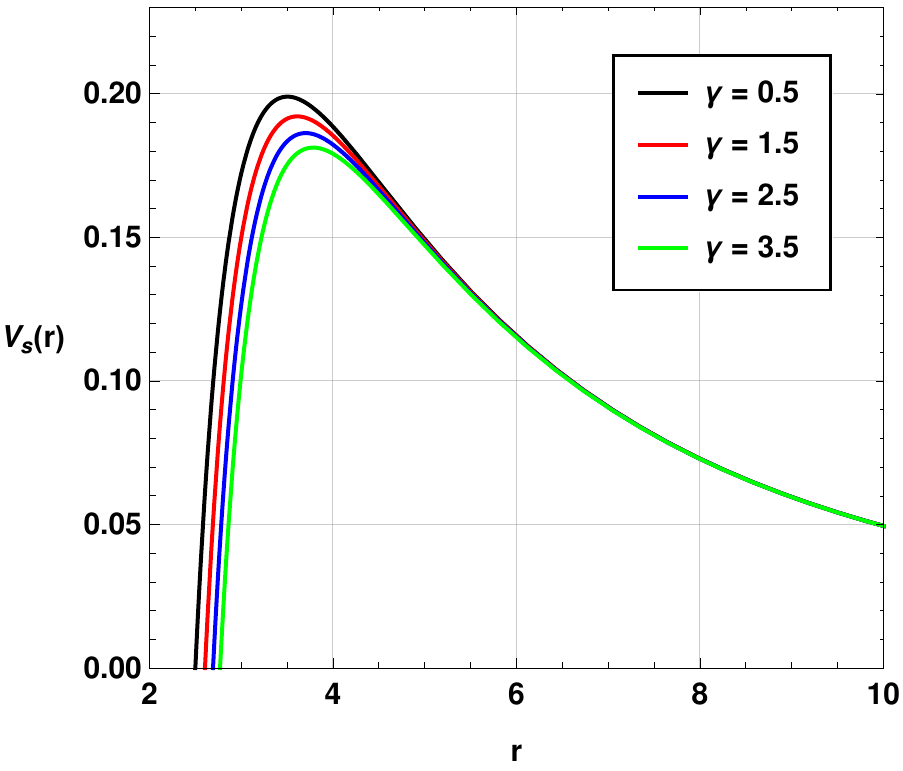}
    }
    \caption{Behavior of scalar potential $V_{s}(r)$ w.r.t. $r$ with $\om = -\frac{41}{10 \pi}$ and $M= G_0 = \hbar = 1.$ On the first panel, we have used $\gamma = 0.1$ and on the second panel, $l=2$.}
    \label{V01}
\end{figure}
\begin{figure}[!ht]
    \centering {
    \includegraphics[scale= 0.70]{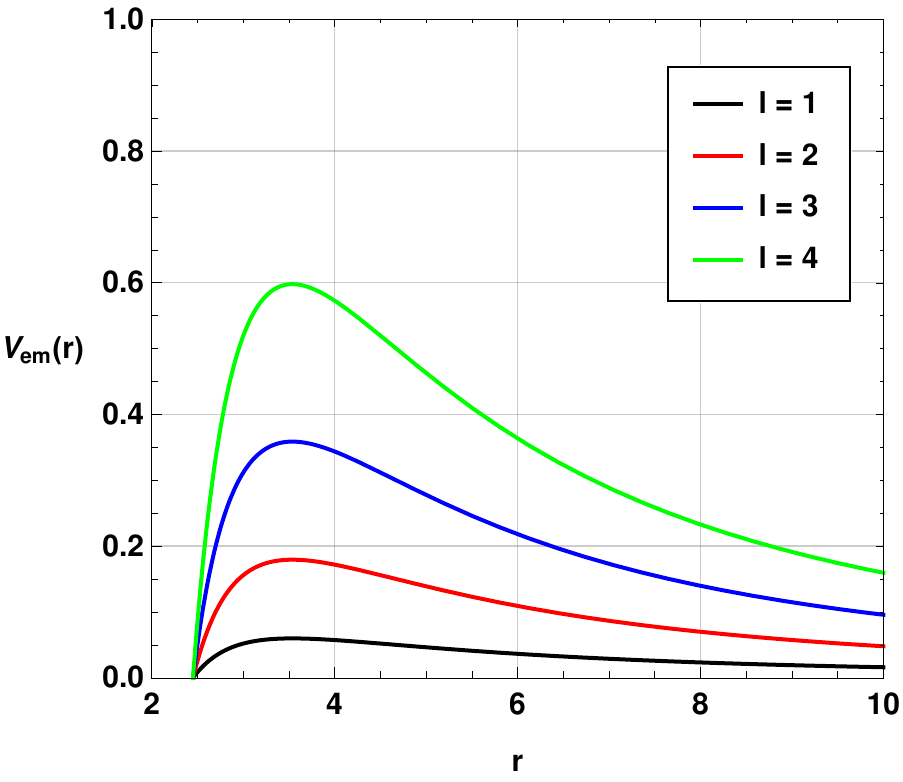} \hspace{1cm}
     \includegraphics[scale=0.710]{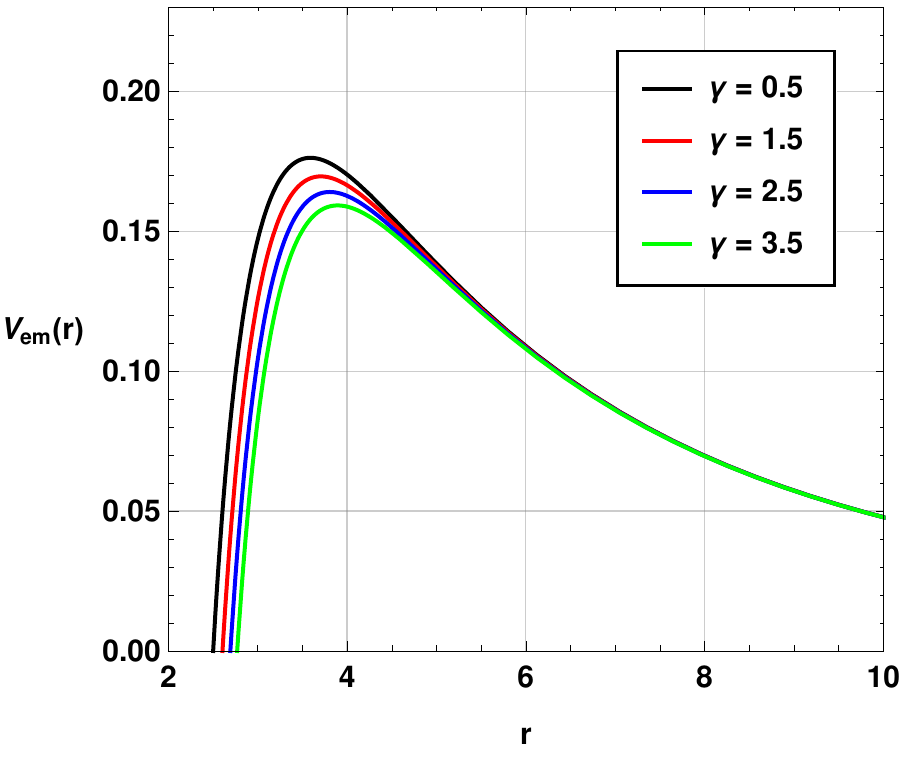}
    }
    \caption{Behavior of electro-magnetic potential $V_{em}(r)$ w.r.t. $r$ with $\om = -\frac{41}{10 \pi}$ and $M= G_0 = \hbar = 1.$ On the first panel, we have used $\gamma = 0.1$ and on the second panel, $l=2$.}
    \label{V02}
\end{figure}

In Fig. \ref{V01}, we have shown the scalar potential for different values of multipole moment $l$ and model parameter $\gamma$. The multipole moment $l$ impacts the potential in a usual way similar to the Schwarzschild black hole. However, the model parameter $\gamma$ has a significantly different impact on the behavior of the potential. One can see that with an increase in the values of the parameter $\gamma$, the peak of the potential shifts towards higher values of $r$. Similar behavior is seen for the electromagnetic potential also (see Fig. \ref{V02}). But in this case, the maximum value of the potential is smaller than the corresponding maximum of the scalar potential.

\subsection{Evolution of perturbations}
Here, we shall discuss the evolution of the perturbation potentials for different values of the model parameters. 
To see the time evolution of the perturbations, we apply the time domain integration formalism as described by Gundlach \cite{Gundlach:1993tn}. To achieve this, we define the variables $\psi(r_*, t) = \psi(i \Delta r_*, j \Delta t) = \psi_{i,j}$ and $V(r(r_*)) = V(r_*, t) = V_{i,j}$. The scalar Klein-Gordon equation can then be expressed as:
\begin{equation}
\dfrac{\psi_{i+1,j} - 2\psi_{i,j} + \psi_{i-1,j}}{\Delta r_*^2} - \dfrac{\psi_{i,j+1} - 2\psi_{i,j} + \psi_{i,j-1}}{\Delta t^2} - V_i\psi_{i,j} = 0.
\end{equation}

To initiate the simulation, we set the initial conditions for $\psi(r_*,t)$ as $\psi(r_*,t) = \exp \left[ -\dfrac{(r_*-k_1)^2}{2\sigma^2} \right]$ with $\psi(r_*,t)\vert_{t<0} = 0$ (where $k_1$ and $\sigma$ are the median and width of the initial wave-packet). We then obtain the time evolution of the scalar field by iterative calculations:
\begin{equation}
\psi_{i,j+1} = -\,\psi_{i, j-1} + \left( \dfrac{\Delta t}{\Delta r_*} \right)^2 (\psi_{i+1, j + \psi_{i-1, j}}) + \left( 2-2\left( \dfrac{\Delta t}{\Delta r_*} \right)^2 - V_i \Delta t^2 \right) \psi_{i,j}.
\end{equation}

By selecting a fixed value of $\frac{\Delta t}{\Delta r_*}$ and utilizing the above iteration scheme, we can obtain the profile of $\psi$ with respect to time $t$. However, it is crucial to ensure that $\frac{\Delta t}{\Delta r_*} < 1$ satisfies the Von Neumann stability condition during the numerical process \cite{Gogoi:2022wyv}.
\begin{figure}[thb!]
    \centering {
    \includegraphics[scale= 0.70]{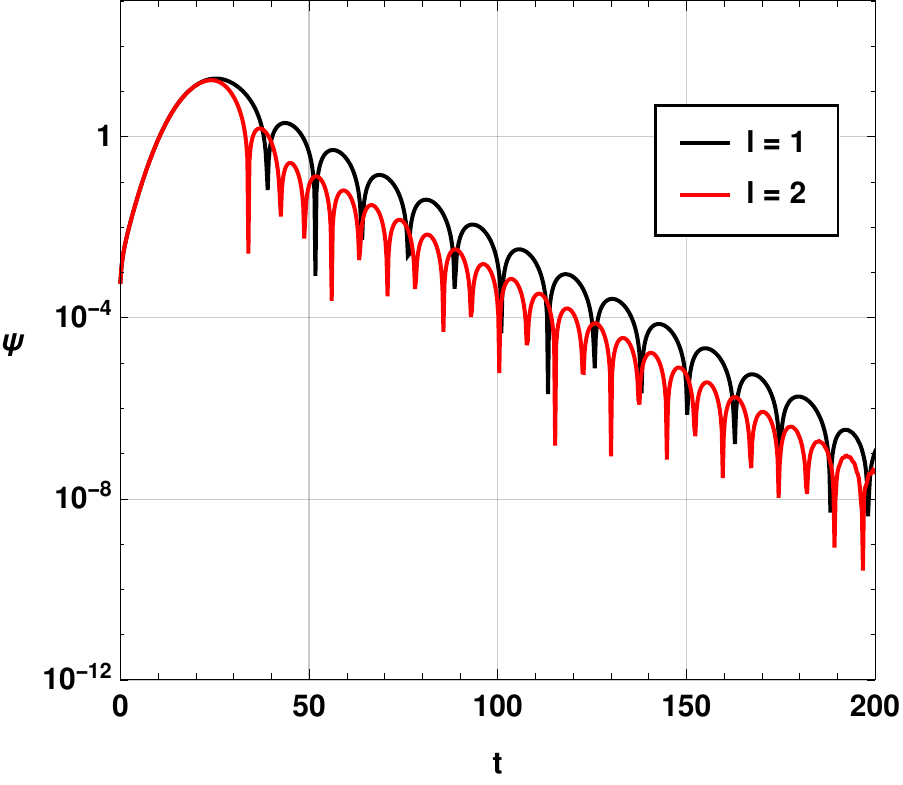} \hspace{1cm}
     \includegraphics[scale=0.70]{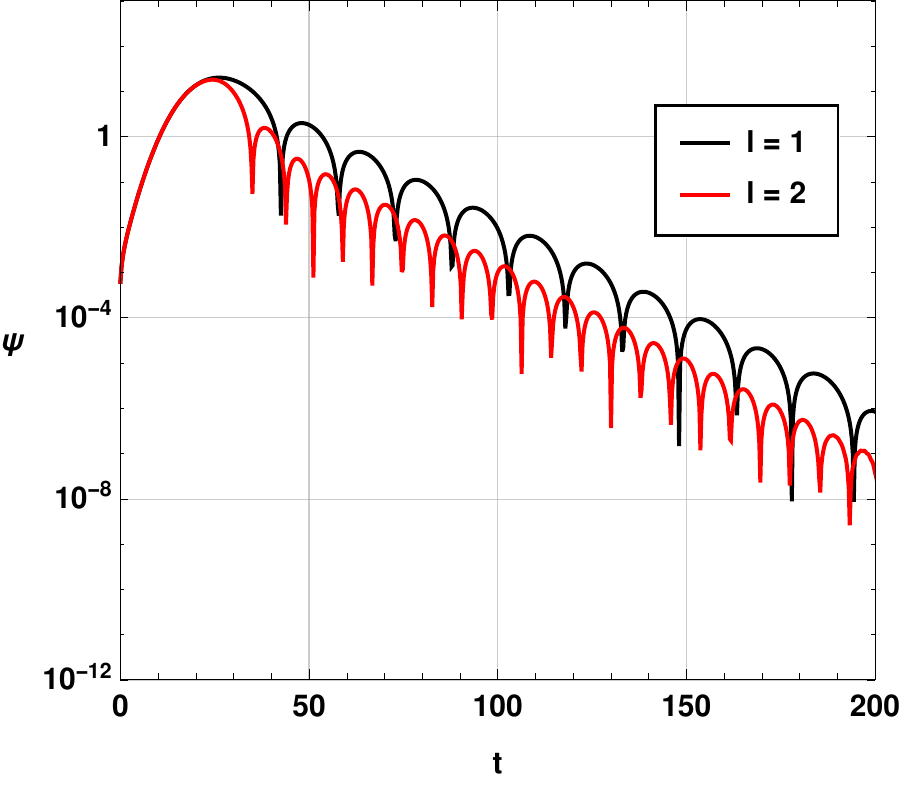}
    }
    \caption{Time domain profiles for scalar (on left) and electromagnetic (on right) perturbations for $\gamma = 1.1$ and $\om = -\frac{41}{10 \pi}$ with $M= G_0 = \hbar = 1.$}
    \label{td01}
\end{figure}
\begin{figure}[bth]
    \centering {
    \includegraphics[scale= 0.70]{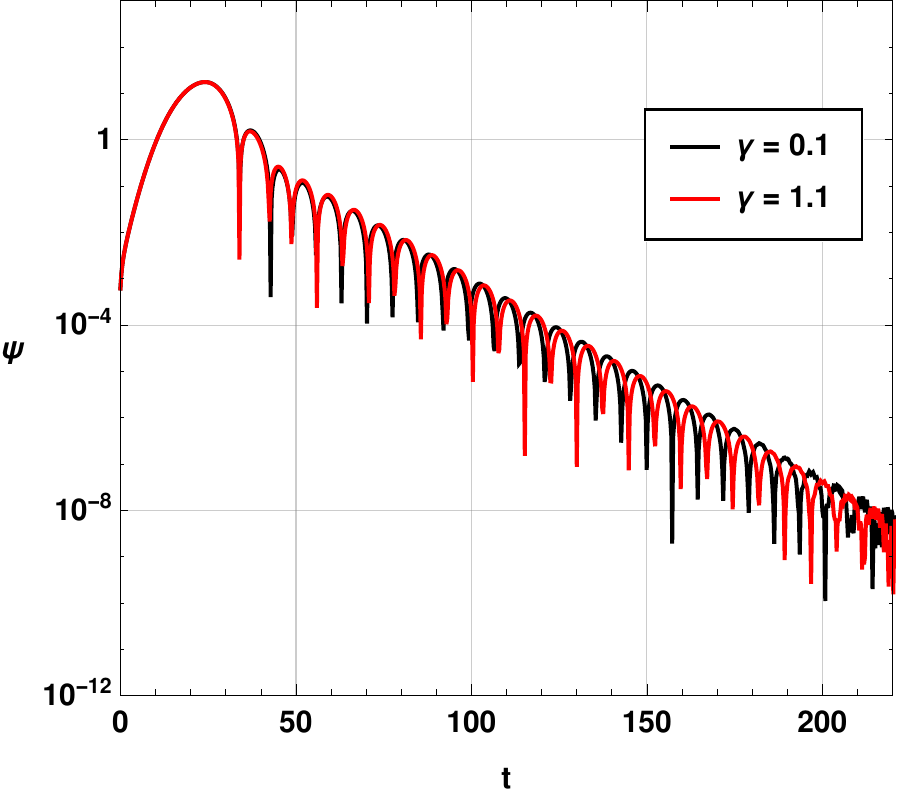} \hspace{1cm}
     \includegraphics[scale=0.70]{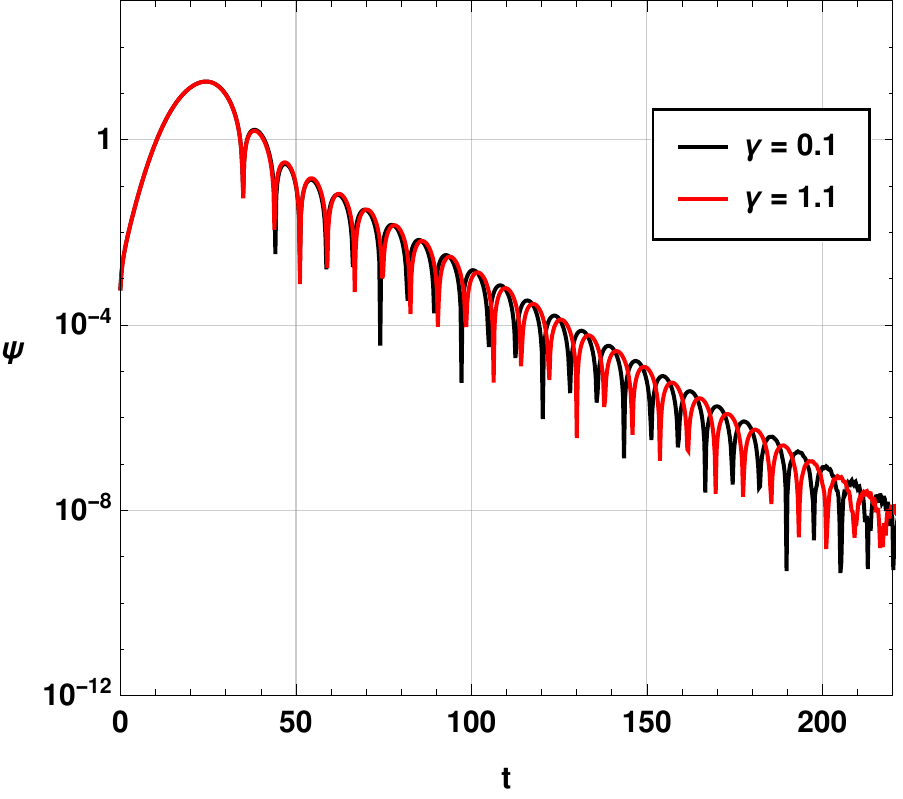}
    }
    \caption{Time domain profiles for scalar (on left) and electromagnetic (on right) perturbations for $l = 2$ and $\om = -\frac{41}{10 \pi}$ with $M= G_0 = \hbar = 1.$}
    \label{td02}
\end{figure}

The time domain profiles for the black hole solution are shown in Fig. \ref{td01} and \ref{td02}. In Fig. \ref{td01}, we have shown the time domain profiles with $l=1$ and $l=2$ for both scalar and electromagnetic perturbations. One can see that the time domain profiles for both perturbations are not identical. In the case of scalar perturbation, the oscillation frequency seems to be higher and the decay rate is also comparatively higher. In both cases, with an increase in the multipole moment $l$, the oscillation frequency increases significantly. However, the variation in decay rate is very small. In Fig. \ref{td02}, we have plotted the time domain profiles for both scalar and electromagnetic perturbations with different values of model parameter $\gamma$. It is interesting to note that the evolution profiles do not show noticeable variations in the initial phase i.e. for a small value of time $t$. However, late-time profiles show a noticeable difference in the oscillation frequencies, and the damping rate is also slightly affected by the variation of the model parameter $\gamma$. To get a more clear idea about the oscillation frequency and damping rate, in the next subsection we shall use the Pad\'e averaged WKB approximation method to calculate the associated quasinormal modes.

\subsection{Quasinormal modes using Pad\'e averaged WKB approximation method}
In this study, we employed the sixth-order Pad'e averaged Wentzel-Kramers-Brillouin (WKB) approximation technique. This method enabled us to calculate the oscillation frequency $\omega$ of gravitational waves (GWs) using the expression below by utilizing the sixth-order WKB method:
\begin{equation}
\omega = \sqrt{-\, i \left[ (n + 1/2) + \sum_{k=2}^6 \bar{\Lambda}_k \right] \sqrt{-2 V_0''} + V_0},
\end{equation}

In this particular context, the variable $n$ refers to overtone numbers and takes on integer values, including $0$, $1$, $2$, and so on. The value of $V_0$ is obtained by evaluating the potential function $V$ at the position $r_{max}$, where the potential reaches its maximum value. At this point, the derivative of $V$ with respect to $r$, i.e. $dV/dr$, becomes zero.

The second derivative of the potential function $V$ with respect to $r$, evaluated at the same position $r_{max}$, is represented as $V_0''$. Additionally, we incorporated supplementary correction terms, which are identified as $\bar{\Lambda}_k$. These correction terms are explicitly defined in Ref.s \cite{Schutz:1985km,Iyer:1986np,Konoplya:2003ii,Matyjasek:2019eeu}. It is important to note that in addition to utilizing the Pad\'e averaging procedure, these correction terms improve the overall accuracy of the calculations.
\begin{table}[t]
\caption{Quasinormal modes of the black hole with $n= 0$, $M=G=\bar{h}=1$, $\om = - \dfrac{41}{10\pi}$ and $\gamma = 0.5$ for the massless scalar perturbation.}
\label{QNMtab01}
\begin{center}
{\small 
\begin{tabular}{|cccc|}
\hline
\;\;\;\;$l$ & \;\;\;\; Pad\'e averaged WKB\;\;\;\;
& \;\;$\vartriangle_{rms}$\;\;\;\; & \;\;$\Delta_6$\;\; \\ \hline
 \;\;$l=1$ & $0.258266\, -0.101353 i$ & $0.0000188516$ & $0.0000888815$\;\; \\
 \;\;$l=2$ & $0.429179\, -0.100341 i$ & $3.631572\times 10^{-6}$ & $0.0000188116$ \;\;\\
 \;\;$l=3$ & $0.600393\, -0.100027 i$ & $4.187839\times10^{-7}$ & $4.083773\times 10^{-6}$\;\; \\
 \;\;$l=4$ & $0.771688\, -0.0998901 i$ & $8.016401\times 10^{-8}$ & $1.248744\times10^{-6}$\;\; \\
 \;\;$l=5$ & $0.94302\, -0.0998188 i$ & $2.186563\times10^{-8}$ & $4.807419 \times 10^{-7}$ \;\;\\
 \;\;$l=6$ & $1.11437\, -0.099777 i$ & $7.572504\times10^{-9}$ & $2.185179\times10^{-7}$ \;\;\\
 \;\;$l=7$ & $1.28574\, -0.0997505 i$ & $3.093985\times 10^{-9}$ & $1.117109\times10^{-7}$ \;\;\\
 \;\;$l=8$ & $1.45711\, -0.0997326 i$ & $1.425313\times10^{-9}$ & $6.221017\times10^{-8}$ \;\;\\ \hline
\end{tabular}
}
\end{center}
\end{table}
\begin{table}[tb]
\caption{Quasinormal modes of the black hole with $n= 0$, $M=G=\bar{h}=1$, $\om = - \dfrac{41}{10\pi}$ and $\gamma = 0.5$ for the electromagnetic perturbation.}
\label{QNMtab02}
\begin{center}
{\small 
\begin{tabular}{|cccc|}
\hline
\;\;\;\;$l$ & \;\;\;\; Pad\'e averaged WKB\;\;\;\;
& \;\;$\vartriangle_{rms}$\;\;\;\; & \;\;$\Delta_6$\;\; \\ \hline
 \;\; $l=1$ & $0.212444\, -0.0937687 i$ & $0.0000310159$ & $0.000199345$\;\; \\
  \;\;$l=2$ & $0.402615\, -0.0979465 i$ & $1.755087\times 10^{-6}$ & $0.0000192351$\;\; \\
  \;\;$l=3$ & $0.581623\, -0.0988517 i$ & $2.492161\times10^{-7}$ & $3.976127\times10^{-6}$ \;\;\\
  \;\;$l=4$ & $0.757157\, -0.0991906 i$ & $5.686693 \times 10^{-8}$ & $1.233163\times10^{-6}$ \;\;\\
  \;\;$l=5$ & $0.931159\, -0.0993542 i$ & $1.733547\times10^{-8}$ & $4.852693\times10^{-7}$ \;\;\\
  \;\;$l=6$ & $1.10435\, -0.0994459 i$ & $6.423011\times10^{-9}$ & $2.232793\times10^{-7}$\;\; \\
  \;\;$l=7$ & $1.27706\, -0.0995025 i$ & $2.738247\times10^{-9}$ & $1.148550\times10^{-7}$ \;\;\\
  \;\;$l=8$ & $1.44946\, -0.0995400 i$ & $1.297526\times10^{-9}$ & $6.416326\times10^{-8}$ \;\;\\ \hline
\end{tabular}
}
\end{center}
\end{table}

In Table \ref{QNMtab01}, we have listed the fundamental quasinormal modes for the massless scalar perturbation for different values of the multipole moment $l$. To obtain these quasinormal modes, we have implemented the above-mentioned Pad\'e averaged 6th-order WKB approximation method. The third column in the table represents rms error associated with the quasinormal modes while the fourth column represents an error term associated with the WKB method defined by \cite{Konoplya:2003ii} 
\begin{equation}
\Delta_6 = \dfrac{\vline \; \omega_7 - \omega_5 \; \vline}{2},
\end{equation}
The variables $\omega_7$ and $\omega_5$ denote the quasinormal modes that were calculated using the $7$th and $5$th order Pad\'e averaged WKB method, respectively. One can see that with an increase in the multipole number $l$, the oscillation frequency increases and the damping rate decreases. From $\vartriangle_{rms}$ and $\Delta_6$, it is clear that the error associated with quasinormal modes with higher $l$ values is significantly small. It is basically due to the property of the WKB approximation method \cite{Gogoi:2022ove, Konoplya:2003ii, Gogoi:2023kjt} which states that for $l>>n$ the accuracy of the WKB method increases significantly.

\begin{figure}[htb!]
    \centering {
    \includegraphics[scale= 0.50]{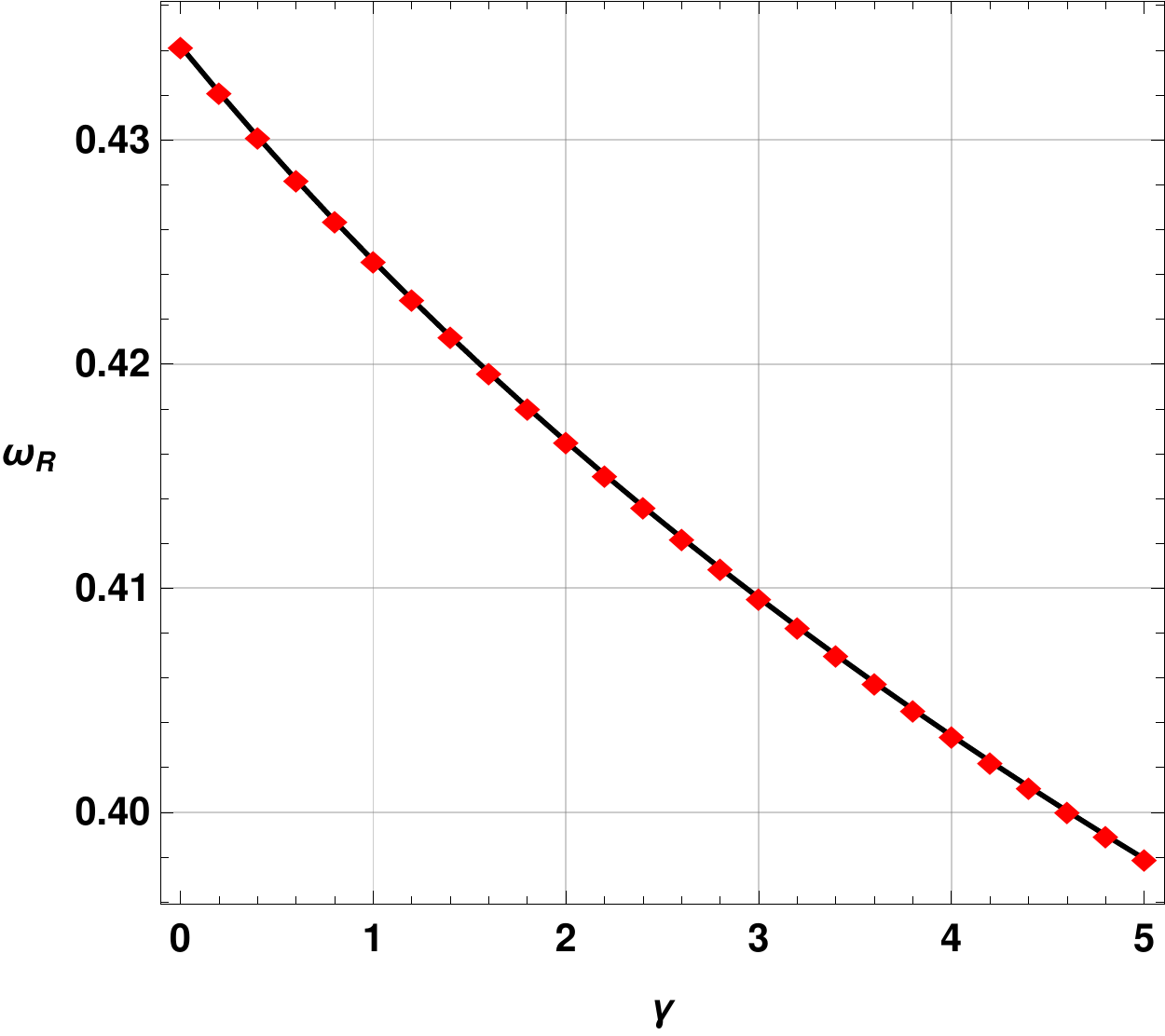} \hspace{1cm}
     \includegraphics[scale=0.510]{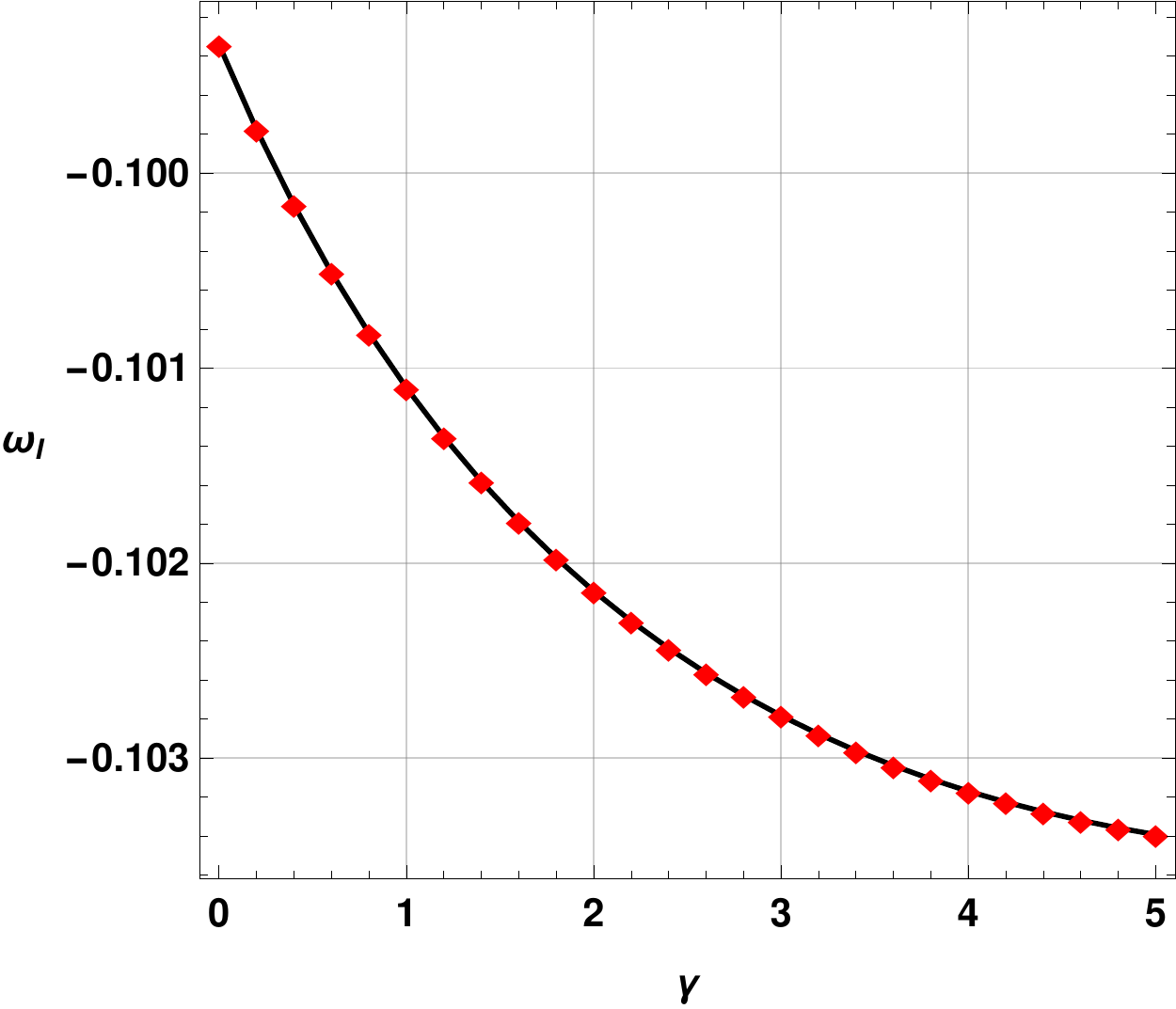}
    }
    \caption{Behavior of scalar quasinormal modes with $n= 0$, $l=2$, $M=G=\bar{h}=1$ and $\om = - \dfrac{41}{10\pi}$.}
    \label{QNMs01}
\end{figure}

\begin{figure}[bth!]
    \centering {
    \includegraphics[scale= 0.50]{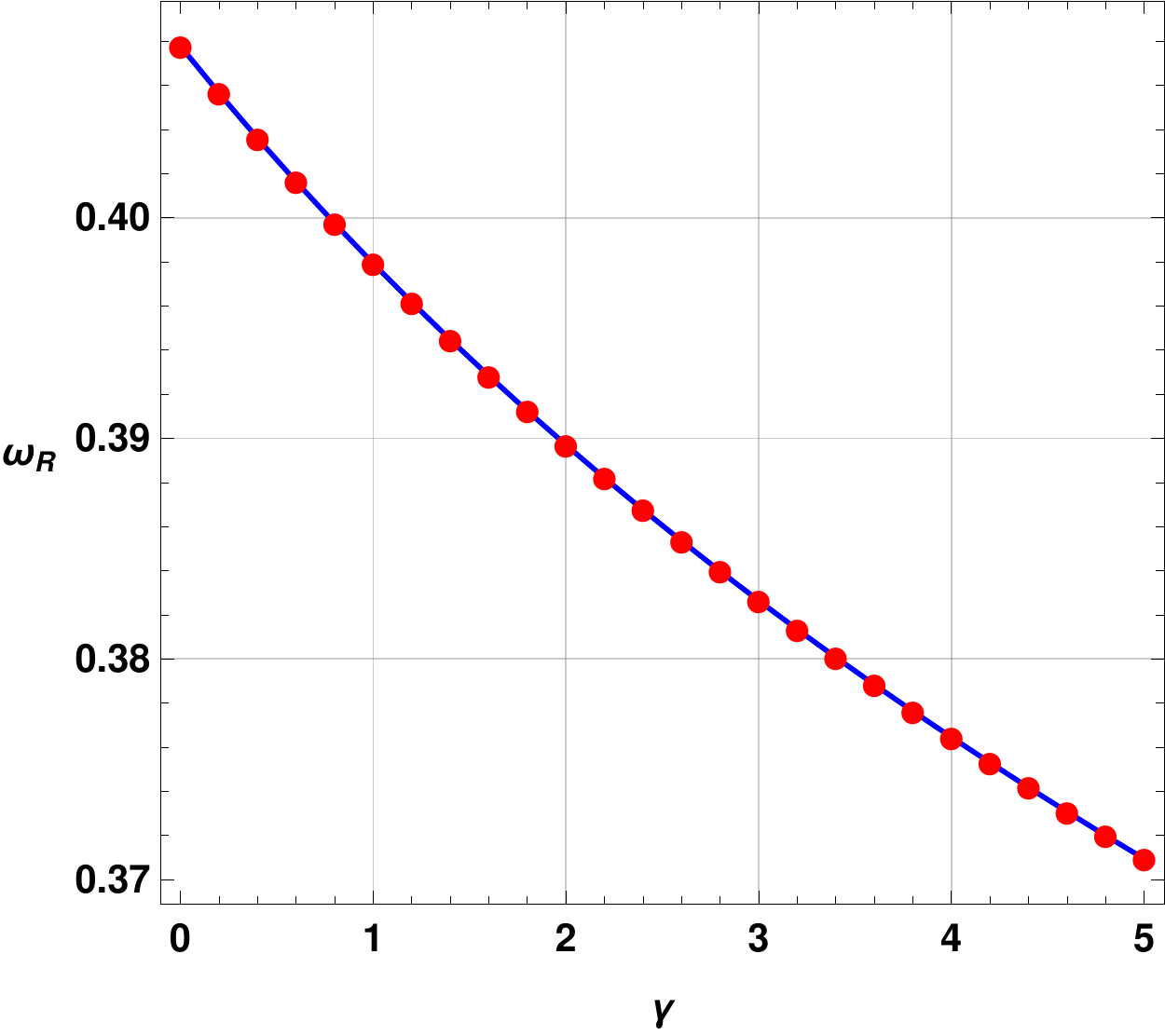} \hspace{1cm}
     \includegraphics[scale=0.510]{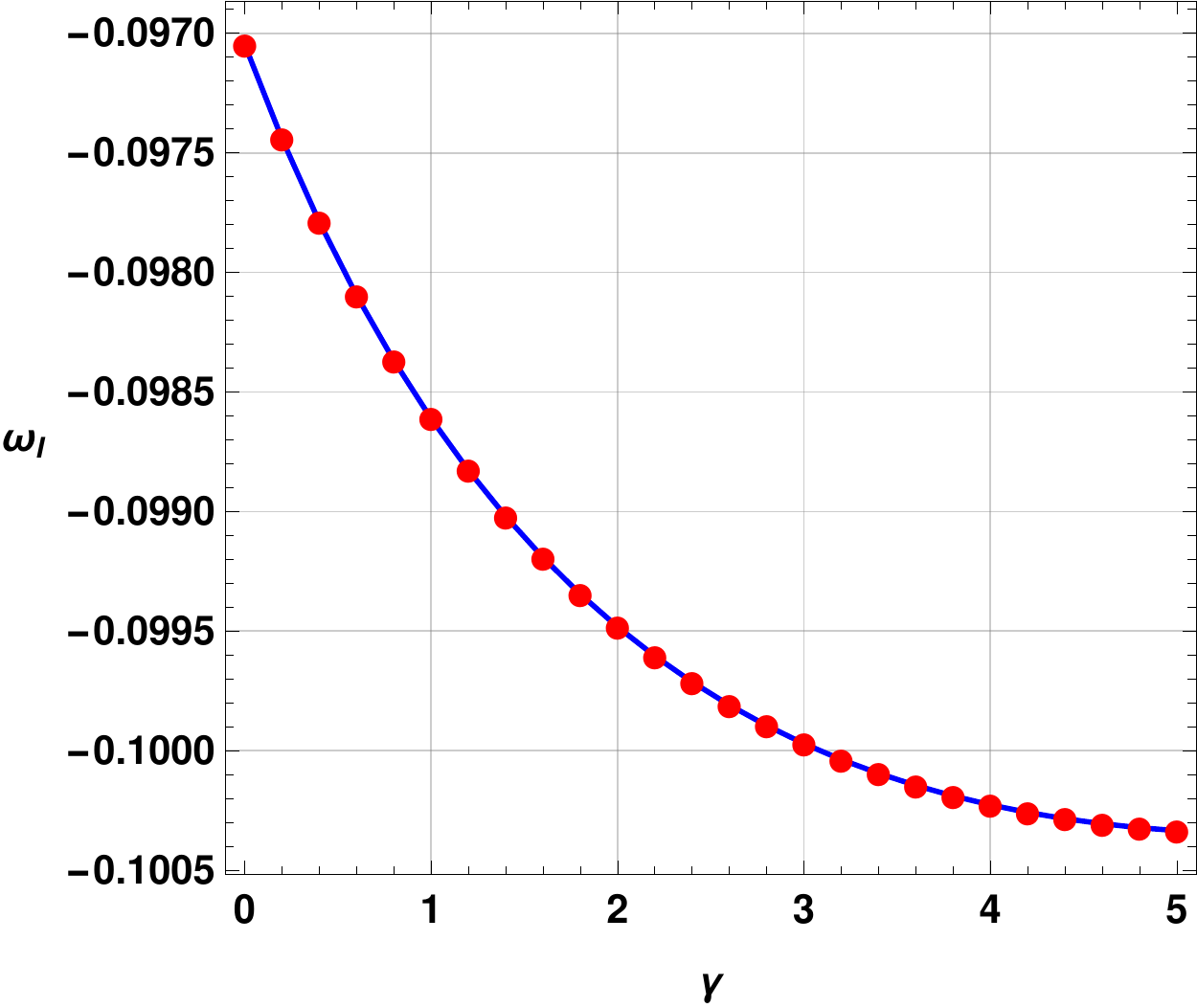}
    }
    \caption{Behavior of electromagnetic quasinormal modes with $n= 0$, $l=2$, $M=G=\bar{h}=1$ and $\om = - \dfrac{41}{10\pi}$.}
    \label{QNMs02}
\end{figure}

To understand the behavior of the scalar quasinormal modes with respect to the model parameter $\gamma$, we have plotted the real and imaginary quasinormal modes in Fig. \ref{QNMs01}. One can see that the real quasinormal modes decrease drastically with an increase in the value of $\gamma$. On the other hand, with an increase in the parameter $\gamma$, the damping rate of gravitational waves increases non-linearly. However, in comparison to the real modes, the impact of $\gamma$ on the damping rate is less significant. We have plotted the real and imaginary quasinormal modes for electromagnetic perturbation in Fig. \ref{QNMs02}. It is clear from the figure that the variation of the quasinormal modes for both cases is similar. However, in the case of electromagnetic perturbation, the quasinormal modes are smaller than those found in the case of scalar perturbation.

The behavior of quasinormal modes is consistent with another study \cite{Anacleto:2021qoe} where the impacts of the GUP parameters on the quasinormal modes have been investigated. However, it should be mentioned that in this Ref. \cite{Anacleto:2021qoe}, the authors considered two deformation parameters having opposite impacts on the quasinormal mode spectrum. The first deformation parameter used in this study resembles the behavior of $\gamma$. In another study \cite{Gogoi:2022wyv}, GUP effects on quasinormal modes have been investigated in the bumblebee gravity framework. However, in this case, the first deformation parameter, which is linked with $\gamma$ of this study, has an opposite impact on the real quasinormal modes. This variation is basically due to the nature of the black hole solution and the presence of other relics used in Ref. \cite{Gogoi:2022wyv}.

This investigation implies that the parameter $\gamma$ which is connected to the GUP deformation parameter can have significant impacts on the gravitational wave frequencies coming from a perturbed Schwarzschild black hole spacetime or ring-down phase. A comparatively large value of the parameter $\gamma$ can also increase the decay rate of gravitational waves. These results can be used in the near future to constrain ASG using observational results from quasinormal modes. One may note that the current gravitational wave detectors may not be able to detect quasinormal modes from black holes with suitable accuracy and hence we might need to wait for LISA to have a more clear and convincing result \cite{Ferrari:2007dd, Gogoi:2022wyv}.

\section{Conclusion} \label{Conc}
In a recent study \cite{Lambiase:2022xde}, intriguing connections were established between the deformation parameter $\beta$ of the generalized uncertainty principle (GUP) and the two free parameters $\omega$ and $\gamma$ of the running Newtonian coupling constant in the Asymptotic Safe gravity (ASG) program. This study prompted us to examine the shadow and quasinormal modes of black holes. Our investigation demonstrates that the approach in \cite{Lambiase:2022xde} offers a valuable framework for exploring the interplay between GUP and quantum gravity. Additionally, our findings affirm the consistency of ASG and GUP, while providing fresh insights into the nature of black holes and their detectable signatures. Ultimately, our work has significant implications for future research in quantum gravity, which we explore in this paper. To do so, first, we have determined the values of $\gamma$ and $\om$ parameters using the EHT observations of the shadow diameter of Sgr. A* and M87*. Our study presents a constraint plot using the shadow of a black hole with a fixed $\tilde{\omega}$ value. Our results, presented in Fig. \ref{shacons} (black solid lines), offer bounds for $\gamma$, which in turn allows us to determine relevant values for $\beta$. Furthermore, our plot provides a visual representation of how the shadow radius behaves as $\gamma$ (or $\beta$) varies. By considering the string theory for GUP and asymptotically safe gravity, we can further constrain $\tilde{\omega}$, as shown in Fig. \ref{shacons} (blue solid lines), with the corresponding bounds for $\tilde{\omega}$ and its value of $\beta$ presented in Table \ref{tab3}.


Next, our investigation indicates that the GUP deformation parameter's connection to the $\gamma$ parameter can have significant implications for gravitational wave frequencies originating from perturbed Schwarzschild black hole spacetime or ring-down phases. Specifically, a higher $\gamma$ value can result in increased gravitational wave decay rates. These findings have the potential to aid in constraining ASG through the utilization of observational data obtained from quasinormal modes in the future. However, it is essential to note that the current precision of gravitational wave detectors may not allow for accurate detection of quasinormal modes from black holes, necessitating the need for the LISA to provide more precise and convincing results \cite{Ferrari:2007dd, Gogoi:2022wyv}.

\acknowledgements
GL, A\"O and RP to acknowledge networking support by the COST Action CA18108.
GL thanks INFN for financial support.

\bibliography{references}
\bibliographystyle{apsrev}

\end{document}